\documentclass[traditabstract]{aa} 
\usepackage{graphicx}
\usepackage{amssymb}
\usepackage{subfigure}
\usepackage{txfonts}
\usepackage{natbib}

\begin{document}

   \title{51~Pegasi - a planet-bearing Maunder minimum candidate}

   \author{K.~Poppenh\"ager\inst{1}
      \and
       J.~Robrade\inst{1}
      \and
          J.~H.~M.~M.~Schmitt\inst{1}
      \and
      J.~C.~Hall\inst{2}}
   \institute{Hamburger Sternwarte, University Hamburg,
             Gojenbergsweg 112, 21029 Hamburg, Germany\\
              \email{katja.poppenhaeger@hs.uni-hamburg.de}
    \and
       Lowell Observatory, 1400 West Mars Hill Road, Flagstaff, AZ 86001, USA
             }

   \date{Received July 22, 2009 / Accepted October 7, 2009}

  \abstract
{We observed 51~Peg, the first detected planet-bearing star, in a 55 ks {\em XMM-Newton} pointing and in 5 ks pointings each with {\em Chandra} HRC-I and ACIS-S. The star has a very low count rate in the {\em XMM} observation, but is clearly visible in the {\em Chandra} images due to the detectors' different sensitivity at low X-ray energies. This allows a temperature estimate for 51~Peg's corona of T$\lesssim1\mbox{MK}$; the detected ACIS-S photons can be plausibly explained by emission lines of a very cool plasma near 200~eV. The constantly low X-ray surface flux and the flat-activity profile seen in optical \ion{Ca}{ii} data suggest that 51~Peg is a Maunder minimum star; an activity enhancement due to a Hot Jupiter, as proposed by recent studies, seems to be absent. The star's X-ray fluxes in different instruments are consistent with the exception of the HRC Imager, which might have a larger effective area below 200~eV than given in the calibration.}


\maketitle



\section{Introduction}

The star 51~Peg (GJ~882, HD~217014) shot to fame in 1995 when \cite{mayorqueloz1995} detected an exoplanet in its orbit, the planetary parameters being quite unexpected at that time, because 51~Peg~b is a giant planet, located at only 0.05~AU distance. The star itself is a G5V star 15.4~pc away from the Sun. Its properties are quite similar to the Sun's, since 51~Peg is about 4~Gyr old and its mass, radius and effective temperature are comparable to solar values with $R=1.27R_{\sun}$ \citep{bainesmcalister2008}, $M=1.11M_{\sun}$, $T_{eff}\approx 5790K$ \citep{fuhrmannpfeiffer1997}. However, 51~Peg is a metal-rich star, for which the metallicities given in the literature vary over a wide range of $+0.05\leq [Fe/H]\leq +0.24$, see for example \cite{valentifischer2005}. Enhanced metallicities are a common feature of stars with giant planets \citep{gonzales1997, santosisraelianmayor2001}.

The activity profile of 51~Peg turned out to be unspectacular. In the Mount Wilson program \citep{baliunasdonahuesoon1995}, which monitors the \ion{Ca}{ii} H and K line fluxes of main sequence stars, the star shows a very low and nearly flat chromospheric activity level from 1977 until 1989 and a slight drop in 1990 and 1991. In the Lowell Observatory program \citep{halllockwoodskiff2007}, it also shows low activity and little variability in \ion{Ca}{} fluxes since the beginning of the program in 1994. The star was also observed in a 12.5~ks {\em ROSAT} PSPC pointing in 1992 and detected as a weak X-ray source.

The coronal activity of 51~Peg is of interest not only because the star is similar to the Sun, but also with regard to recent studies \citep{kashyapdrakesaar2008}, which claim stars with close-in giant planets to be more X-ray active than stars with far-out ones.

\section{Observations and data analysis}

We observed 51~Peg on two occasions in 2008. A 55 ks {\em XMM-Newton} was carried out on June 1, 2008, and with {\em Chandra}, we observed 51~Peg for 5 ks each using HRC-I and ACIS-S on December 6, 2008 immediately after each other. The specific observation details are listed in Table \ref{Obslog}.

   \begin{table*}
      \caption[]{{\em XMM} and {\em Chandra} Observation log of 51~Peg}
        \label{Obslog}
    \begin{tabular}{l l l l r}
    \hline\hline
    Instrument & Configuration & ObsID & Obs. time & GTI (s) \\ \hline
    {\em XMM} MOS1 & full frame / thick filter & 0551020901 & 2008-06-01 11:57:03 2008-06-02 03:17:50 & 55000\\
    {\em XMM} MOS2 & full frame / thick filter & 0551020901 &  2008-06-01 11:57:03 2008-06-02 03:17:55 & 55100\\
    {\em XMM} PN & full frame / thick filter & 0551020901 &  2008-06-01 13:09:38 2008-06-02 03:18:10 & 29000\\
    {\em Chandra} ACIS-S & back-illuminated & 10825 & 2008-12-06 11:03:26 2008-12-06 12:26:26 & 4980\\
    {\em Chandra} HRC-I & imaging & 10826 & 2008-12-06 12:44:54 2008-12-06 14:07:54 & 4924 \\ \hline
    \end{tabular}
   \end{table*}

\subsection{{\em XMM-Newton} data analysis}

The {\em XMM-Newton} data were reduced using the Science Analysis System (SAS) version 8.0.0. Standard selection criteria were applied for filtering the data. In the full-time image obtained with the PN detector, the automatic source detection procedure finds a faint X-ray source with 32 excess counts at 51~Peg's nominal position when using the 0.2-1~keV energy band. This choice is motivated by 51~Peg being detected in the 1992 {\em ROSAT} PSPC pointing as a very soft X-ray source. Because of the weak signal, we merged both MOS detectors. In the RGS, no relevant signal was present. The PN observation is affected by proton contamination, therefore we used only time intervals (GTI) where the high-energy background averaged over the detector is below 0.8~cts/s, leading to a PN GTI of 29~ks.

Since spectral fitting results are not very reliable with this low number of counts, we conducted a study in different energy bands instead and investigated the recorded counts within the source region, a radius of 15$\arcsec$ around 51~Peg's nominal position for the PN and MOS instruments. The source region size of 15$\arcsec$ radius was chosen because of the rather broad point spread function, which contains 72\% (68\%) of the photons from a point-like source in the PN (MOS) detector. Background counts were extracted from source-free nearby regions, which are located on the same chip for the MOS detector; for the PN detector, two background regions were chosen, one on the same chip as the source and one on a neighboring chip. Since 51~Peg proved to be a very soft X-ray source in the previous {\em ROSAT} observation, we expect most X-ray photons to be produced from the \ion{O}{vii} triplet or lines with even lower energies, such as \ion{N}{vii/vi} and \ion{C}{vi/v}. We therefore specified three energy bands for our analysis, concentrating on a band around the \ion{O}{vii} triplet ($\approx570$~eV); the detected photons are given in Table~\ref{XMMcounts}. Since the EPIC detectors have energy resolutions of FWHM$\approx$100~eV, we adopted this as minimum bandwidth.

   \begin{table}
\setlength\tabcolsep{5pt}
      \caption[]{Photons of 51~Peg in {\em XMM} and {\em Chandra}; see text for details.}
        \label{XMMcounts}
    \begin{tabular}{l l l l l  l l l l}
    \hline\hline
    Energy & \multicolumn{2}{l}{PN} & \multicolumn{2}{l}{MOS1+2} & \multicolumn{2}{l}{ACIS-S} & \multicolumn{2}{l}{HRC}\\ \vspace{0.1cm}
    (keV)& s & b & s & b & s & b & s & b\\ \hline  \vspace{0.1cm}
    0.15-0.2   & ... & ... & ... & ... & $1^{+2.4}_{-0.4}$ & 0 & ... &...\\  \vspace{0.1cm}
    0.2-0.45   & $17^{+5.0}_{-3.0}$ & 9.1  & $7^{+3.6}_{-1.6}$ & 4.9 & $6^{+3.6}_{-1.6}$ & 0 &...&... \\  \vspace{0.1cm}
    0.45-0.65 & $8^{+3.9}_{-1.9}$ & 3.5 & $6^{+3.5}_{-1.5}$ & 3.2 & $0^{+2}_{-0}$ & 0 & ... & ...\\  \vspace{0.1cm}
    0.65-2.0   & $10^{+4.3}_{-1.5}$ & 10.4 & $21^{+7.2}_{-0.8}$ & 23.2 & $0^{+2}_{-0}$ & 0 & ...&...\\ \vspace{0.1cm}
    0.15-2.0   & ... & ...& ... & .... & ...& ...& $21^{+5.7}_{-3.7}$ & 0.6 \\
    \hline
    \end{tabular}
   \end{table}
In the source region we then count the number of photons recorded in the various energy bands and detectors, listed in column ''s''  in Table~\ref{XMMcounts}. Source counts are given with $1\sigma$ confidence limits for low count numbers according to \cite{kraftburrowsnousek1991}; for a detailed discussion, see \cite{ayres2004}. The scaled background counts (denoted ''b'') were taken from areas much larger than the source region, and thus the error on the background is dominated by statistical fluctuations.

\subsection{{\em Chandra} data analysis}

For data reduction of the {\em Chandra} observations we used CIAO 4.1 \citep{ciao} and applied standard selection criteria. The analysis of the data was performed in the 0.15\,--\,1~keV energy band since the back-illuminated ACIS-S chip has nonzero effective area at X-ray energies below 300~eV. For the HRC imager no energy cuts were used since its energy resolution is low. 51~Peg is clearly detected in both instruments.

In the ACIS observation we detect eight photons in the source region, a circle with 1.5$\arcsec$ radius around 51~Peg's nominal position. This radius was chosen to contain 95\% of the soft ($\leq$1~keV) photons from a point-like source. From nearby source-free regions in the 150\,--\,650~eV energy band we expect only 0.03~background counts for this area, therefore we attribute all of the recorded counts to 51~Peg. The spectral resolution of ACIS-S is similar to the one of the EPIC detectors ($\approx$ 100~eV).

In the HRC-I pointing 21~photons were detected in the source region over a background of 0.6~photons scaled to the same area. At any rate, also the HRC clearly detects 51~Peg.

\section{Results}

\begin{figure}
\includegraphics[width=0.5\textwidth]{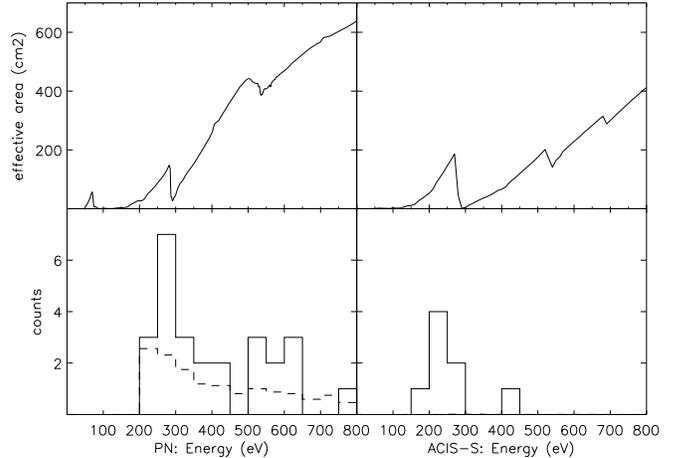}
\caption{{\em Upper panel:} effective areas of {\em XMM} PN and {\em Chandra} ACIS-S at energies below 800~eV. {\em Lower panel:} detect cell counts (solid histogram) of 51~Peg in PN and ACIS-S over the respective backgrounds (dashed; ACIS-S has practically no background).}
\label{spectra}
\end{figure}

\subsection{{\em XMM-Newton} PN and MOS}

51~Peg shows a photon excess in the 0.2\,--\,0.45~keV and the \ion{O}{vii} band (0.45\,--\,0.65~keV) in PN and a very weak excess in the same bands in the merged MOS detectors. The MOS and PN lightcurves show no obvious variability over the whole 55~ks. As shown in Figure~\ref{spectra}, most of PN's excess source photons have energies around 300~eV; another emission feature is present around 570~eV, the energy of the \ion{O}{vii} triplet. Because of {\em XMM}'s moderate intrinsic energy resolution the nominal energies of the detected source photons cannot be regarded as exact values. From the absence of emission features at \ion{O}{viii} energies ($\approx650$~eV) we can conclude that 51~Peg's corona has an average plasma temperature well below 3~MK.

\subsection{{\em Chandra} ACIS-S and HRC-I}

All the recorded counts have energies between 150 and 450~eV and are distributed quite evenly over the observation time, supporting a soft, basically constant X-ray source. Let us now inspect the energies of the ACIS-S photons in detail; the CIAO software assigns a nominal energy to each recorded photon (see Figure~\ref{spectra}). The eight source photons have energies of 170, 206, 211, 212, 256, 227, 291 and 428~eV; they are hence very soft and obviously none of these photons can be attributed to \ion{O}{vii} or even \ion{O}{viii} emission. This supports our hypothesis of a very low plasma temperature evoked by the {\em XMM}~data.

The ACIS-S detector is prone to optical contamination, so we have to check whether the extremely soft events could be induced by optical photons. The threshold for optical contamination in the ACIS-S detector is at $V\approx 7.8$ for stars with an effective temperature between 5000 and 6500~K; a star this bright would cause a bias level shift of one Analog-to-Digital-Unit (ADU) of 3.4~eV during the standard 3.2~s time frame for the central pixel of the source. 51~Peg's visual magnitude is 5.5, so we expect ca. 8~ADUs per time frame. Since the event threshold lies at 20~ADUs, optical contamination can be ruled out as explanation for the detected events.

Also in the HRC the recorded events are distributed evenly throughout the observation time. The intrinsic energy resolution of the HRC detector is low so that little information on the spectral energy distribution can be derived; because the HRC-I observation was carried out immediately after the ACIS-S observation, we assume that the 21 detected HRC-I source photons have similar energies as the photons in the ACIS-S detector.

   \begin{table}[!t]
      \caption[]{X-ray fluxes of 51~Peg with 1$\sigma$ errors, calculated with WebPIMMS using a 1\,MK thermal plasma model.}
        \label{fluxes}
    \begin{tabular}{l l l}
    \hline\hline
    Instrument &  Flux (0.1\,--\,1.0~keV) & $\log L_X$ (0.1\,--\,1.0~keV)  \\
	  & (erg s$^{-1}$ cm$^{-2}$) & (erg s$^{-1}$)\\ \hline \vspace{0.1cm}
    {\em XMM} PN & $1.2^{+1.0}_{-0.5}\times10^{-14}$ & 26.3\,--\,26.8\\ \vspace{0.1cm}
    {\em XMM} MOS1+2 &   $1.1^{+1.4}_{-0.6}\times10^{-14}$ & 26.1\,--\,26.8\\ \vspace{0.1cm}
    {\em Chandra} ACIS-S &  $1.7^{+0.9}_{-0.4}\times10^{-14}$  & 26.5\,--\,26.8\\ \vspace{0.1cm}
    {\em Chandra} HRC-I & $4.2^{+1.1}_{-0.7}\times10^{-14}$  & 27.0\,--\,27.2\\ \vspace{0.1cm}
    {\em ROSAT} PSPC & $2.1^{+0.3}_{-0.3}\times10^{-14}$  & 26.7\,--\,26.8\\
    \hline
    \end{tabular}
   \end{table}

\subsection{Determination of coronal temperature}
To estimate the coronal temperature, we evaluate the temperature-dependent hardness ratios of several energy bands, viz. $HR_{PN}=H_{PN}/S_{PN}$ with $H_{PN}$ and $S_{PN}$ covering 0.45\,--\,0.65~keV and 0.2\,--\,0.45~keV for PN; for ACIS-S, we use $HR_{ACIS\,1}=H_{ACIS\,1}/S_{ACIS\,1}$ and $HR_{ACIS\,2}=H_{ACIS\,2}/S_{ACIS\,2}$ with the energy bands $H_{ACIS\,1}$: 0.25\,--\,0.45~keV, $S_{ACIS\,1}$: 0.15\,--\,0.25~keV, $H_{ACIS\,2}$: 0.45\,--\,0.65~keV and  $S_{ACIS\,2}$: 0.25\,--\,0.45~keV. The energy bands for ACIS-S were chosen to obtain quite an even distribution of source photons.

\begin{figure}
\subfigure[$HR_{ACIS-S\,1}$]
{\includegraphics[width=0.24\textwidth]{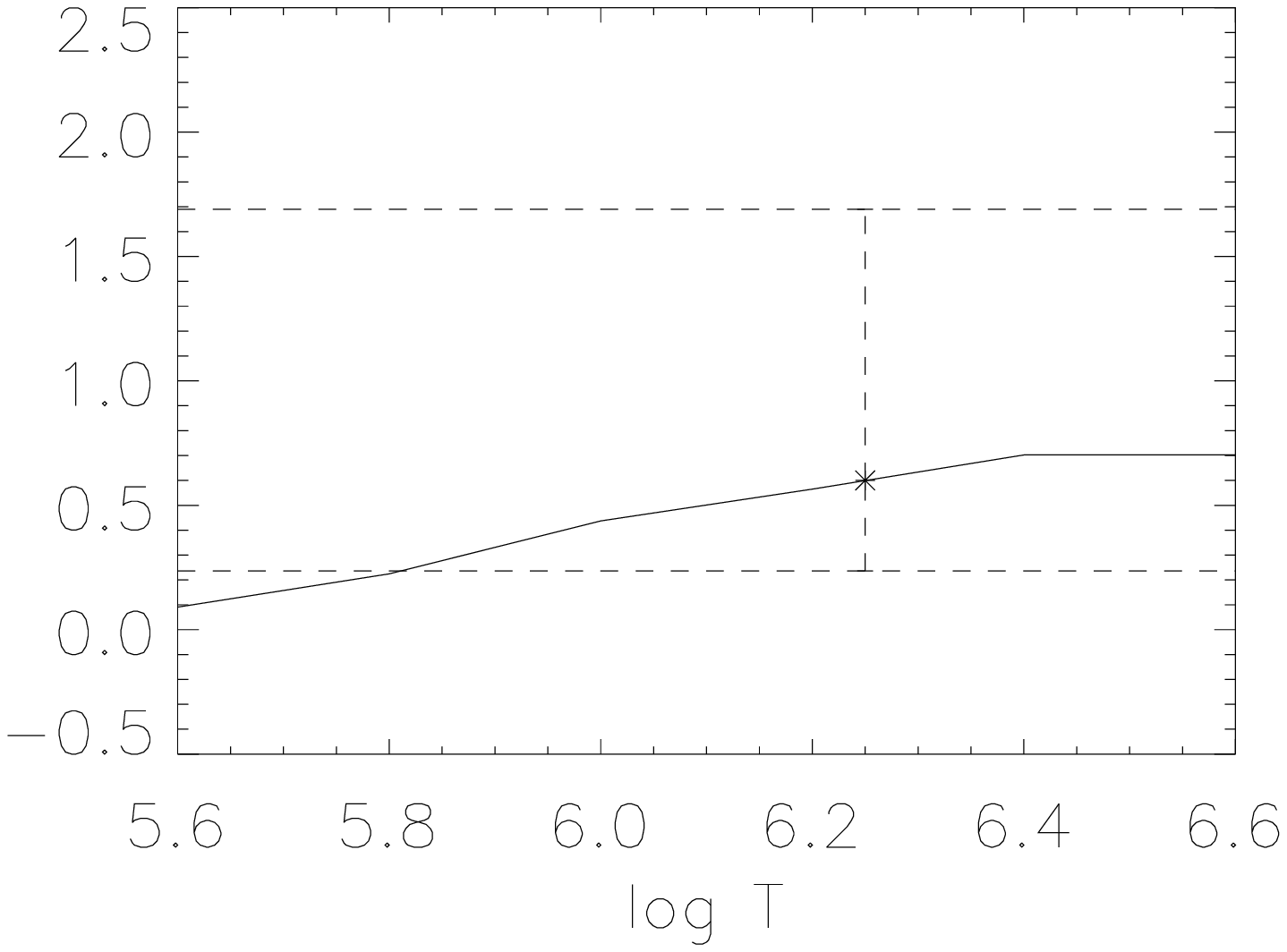}}
\subfigure[$HR_{ACIS-S\,2}$]
{\includegraphics[width=0.24\textwidth]{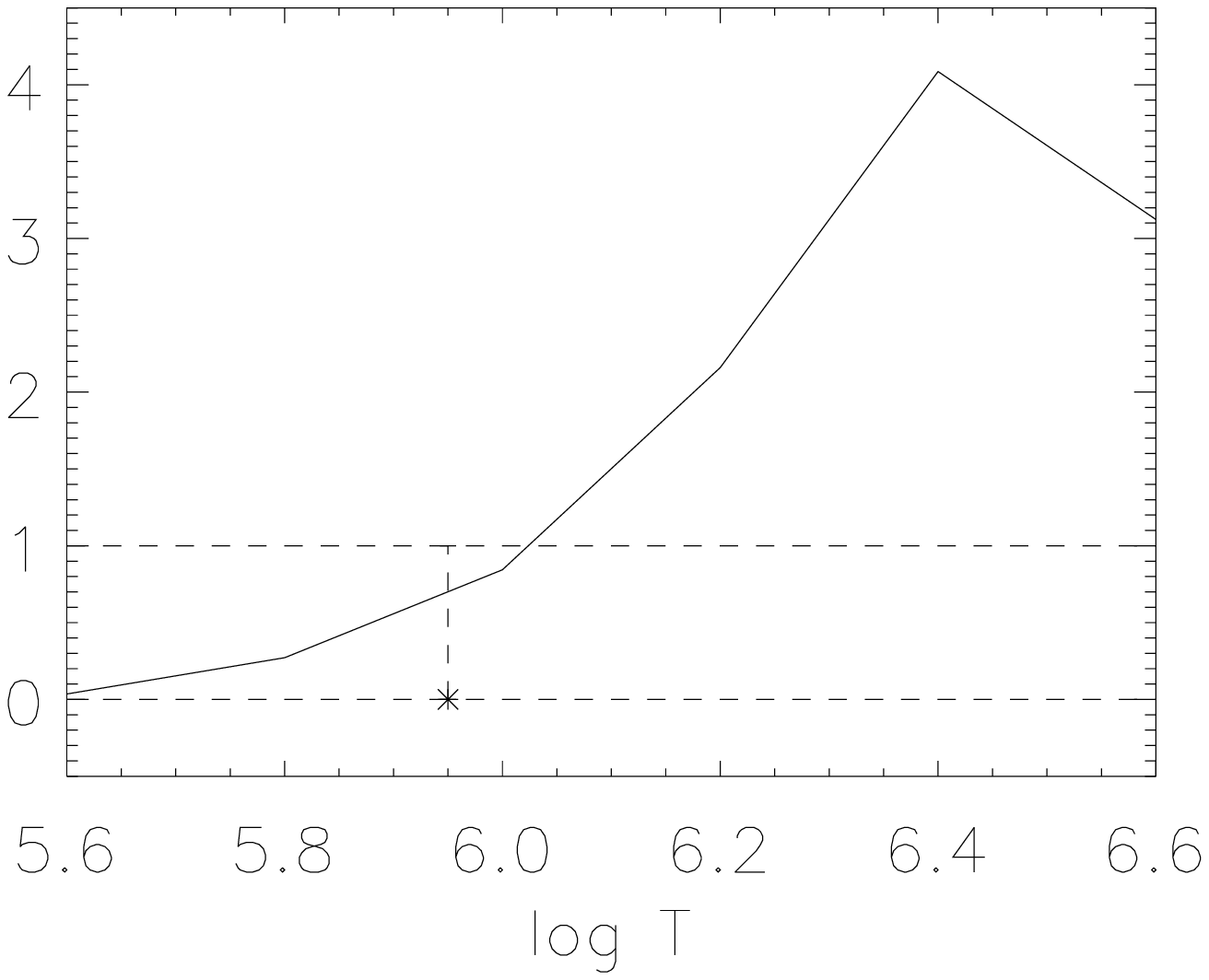}}
\subfigure[$HR_{PN}$]
{\includegraphics[width=0.24\textwidth]{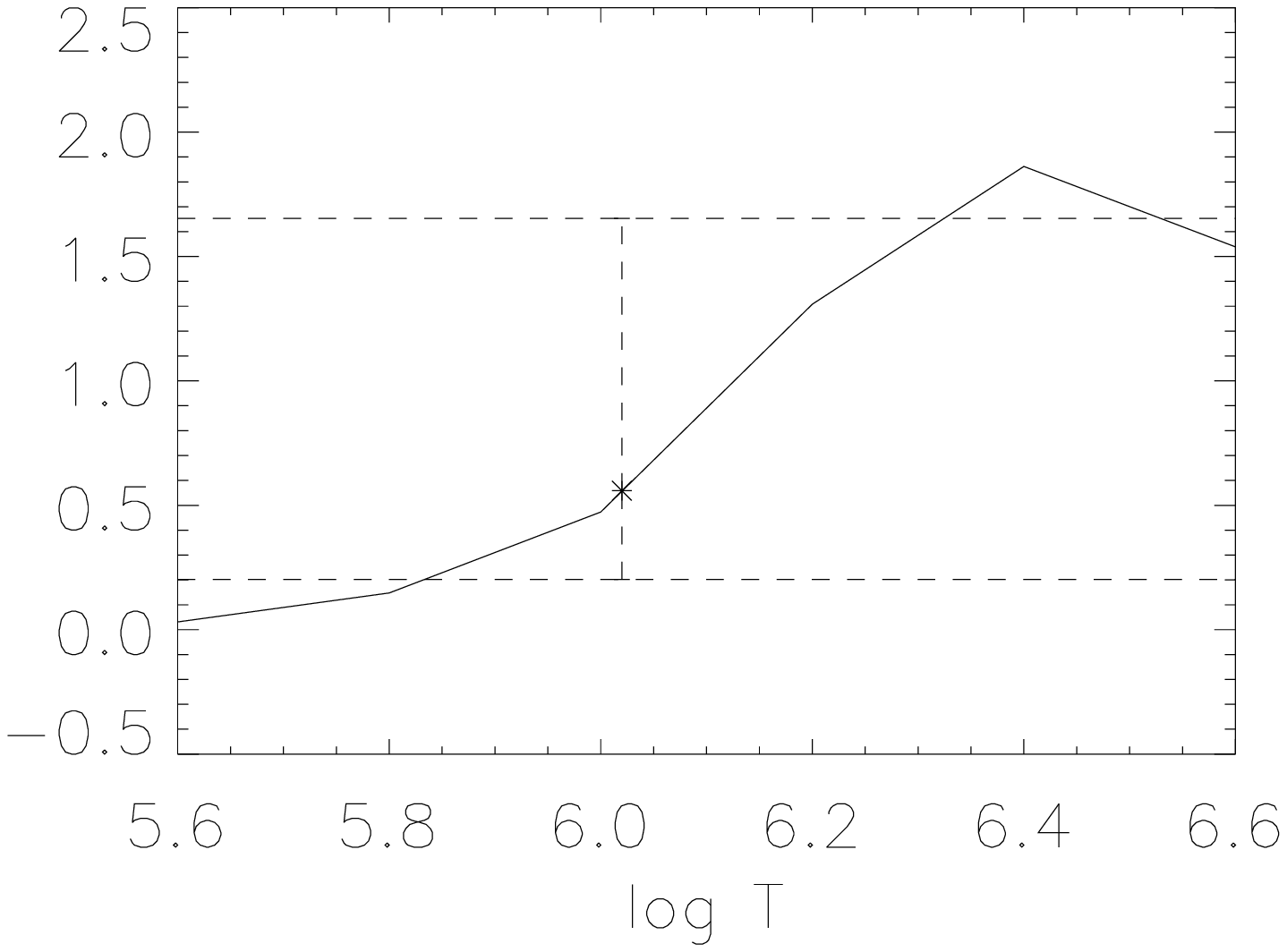}}
\subfigure[$HR_{PSPC}$]
{\includegraphics[width=0.24\textwidth]{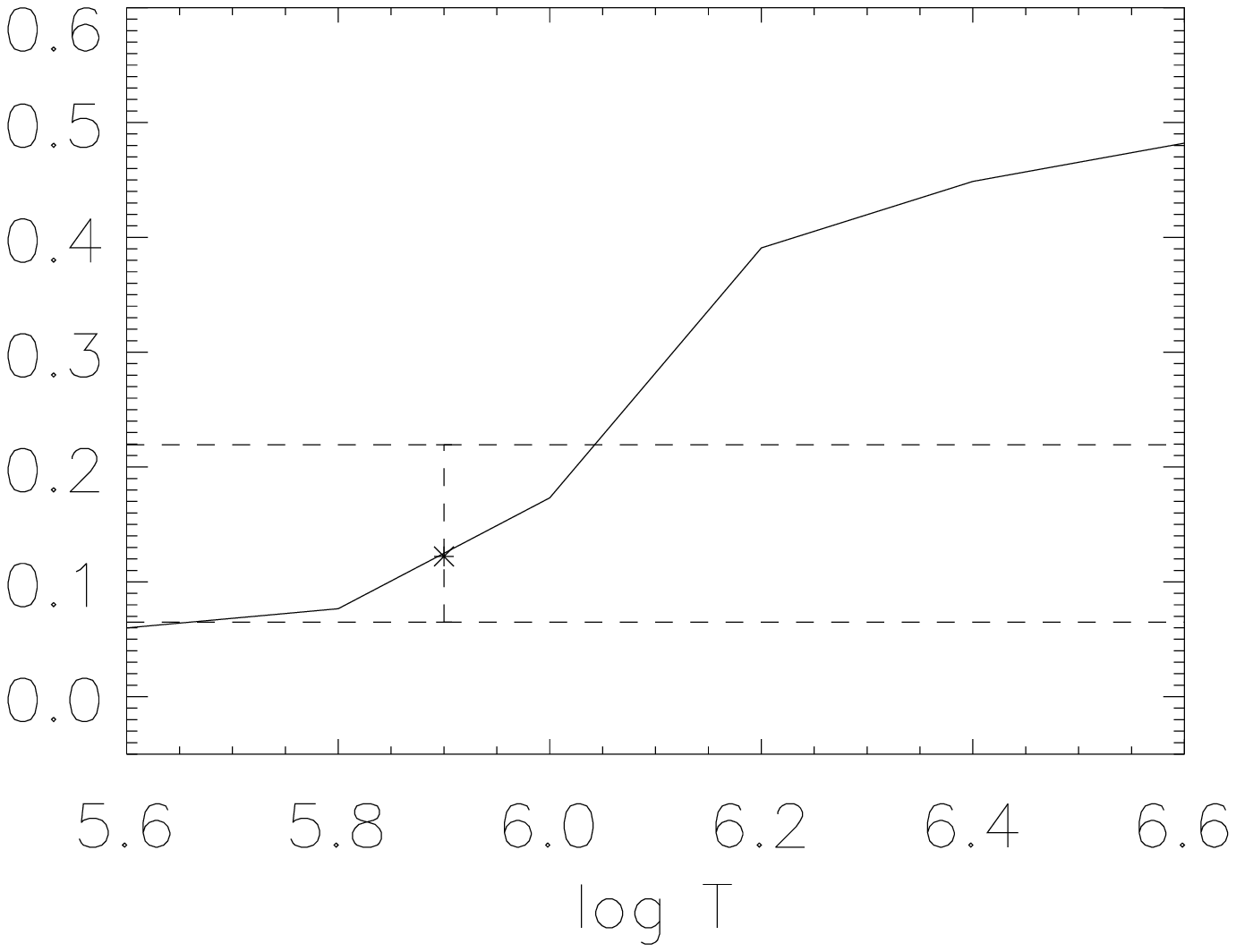}}
\caption{Hardness ratios as a function of temperature (see text for details). Observed ratios are given as crosses with $1\sigma$~error.}
\label{images}
\end{figure}

We derive the temperature-dependence of the hardness ratios with Xspec~v12, using the instrument responses and effective areas as shown in Figure~\ref{spectra} and simulating spectra for a one-temperature thermal plasma model with solar abundance in the temperature range between $\log T=5.6$ and $6.6$ (see Figure~\ref{images}). The possible plasma temperature of 51~Peg is then constrained by the observed hardness ratios for ACIS-S and PN. Assuming $1\sigma$~errors for low count numbers as defined in Table~\ref{XMMcounts}, we find that $HR_{PN}=0.6^{+1.1}_{-0.4}$, $HR_{ACIS\,1}=0.6^{+1.0}_{-0.4}$ and $HR_{ACIS\,2}=0^{+1}_{-0}$. The observed PN ratio yields the temperature limits $5.85 \leq \log T \leq 6.3$. The ACIS-S ratios yield a lower temperature limit from $HR_2$ and an upper limit from $HR_3$: $5.8\leq \log T \leq 6.05$, so that the likely temperature range for 51~Peg's corona is $5.85\leq \log T \leq 6.05$.

Since there are virtually no background photons in ACIS-S, we can use the energies of the eight recorded source photons to validate our temperature constraints by identifying the most likely emission lines of their origin. Strong emission at temperatures near 1~MK comes from the \ion{Si}{} and \ion{S}{} emission line complexes around 200~eV, some strong \ion{Si}{} lines around $\approx$~230~eV, the \ion{C}{v} triplet around 300~eV and the \ion{N}{vi} triplet around 426~eV. These emission lines match well with the observed photons, which can be considered as a rough plausibility check.
So, both the temperature constraints from hardness ratios and the identification of possible emission line complexes point to a plasma temperature of  $T \lesssim 1$~MK. 

\subsection{Comparison with {\em ROSAT} data}

To investigate the long-term X-ray behavior of 51~Peg, we reanalyzed a 12.5~ks {\em ROSAT} PSPC observation from December 1992. Using the conversion factor $ecf=(5.30\times HR_1+8.31)\times 10^{-12}\ \mbox{erg}\ \mbox{count$^{-1}$}\ \mbox{cm$^{-2}$}$ with $HR_1$ being the hardness ratio $HR_1=(H-S)/(H+S)$ (S: 0.1\,--\,0.4~keV, H: 0.5\,--\,2.0~keV) \citep{schmitt1997}, the luminosity derived from the observed count rate of 7~cts/ks is log$L_X=26.75\ \mbox{erg}\ \mbox{s$^{-1}$}$. The observation is split in two parts, interrupted for ca. 70~ks. 51~Peg emits mainly soft X-ray photons below 500~eV ($HR_1\approx -1.0$) and showed a low and constant X-ray activity level without any obvious variability.

We can constrain the coronal temperature of 51~Peg in this observation in the same way as for the {\em XMM} and {\em Chandra} pointings. We use the temperature-dependent hardness ratio $HR_{PSPC}=H_{PSPC}/S_{PSPC}$ with the energy bands $H_{PSPC}$: 0.1\,--\,0.3~keV, $S_{PSPC}$: 0.3\,--\,0.65~keV. The observation yields $HR_{PSPC}=0.12^{+0.10}_{-0.06}$ from which lower and upper limits for the temperature can be derived, namely $5.65\leq \log T \leq 6.05$ (see Figure~\ref{images}). This again leads to a temperatur estimate of $T\lesssim 1$MK.

\subsection{Consistency of measured count rates and fluxes}

The observed count rates in {\em XMM} PN and MOS and {\em Chandra} ACIS-S are, considering the low photon numbers and therefore large statistical fluctuation, in reasonably good agreement. That ACIS-S detects no photons in the \ion{O}{vii} band is no surprise given the shorter exposure in ACIS-S and its smaller effective area in that energy range (see Figure~\ref{spectra}). At energies below 300~eV, we find that the PN and MOS counts numbers are smaller than expected from what we see in ACIS-S. This might be explained by statistical fluctuations, errors in the effective area determination or energy redistribution effects in the CCD detectors (EPIC "low-energy shoulder"). Given these uncertainties for very low energies, we use only the \ion{O}{vii} counts (0.45-0.65~keV) of PN and MOS for our flux calculations and then extrapolate the flux to a common energy range of 0.1\,--\,1~keV for comparison. For the other instruments, we use 0.65~keV as the upper bound of the energy range and their low-energy sensitivity limits as the lower bound (0.15~keV for ACIS-S/HRC, 0.1~keV for PSPC) and then extrapolate to the common energy range.

The fluxes normalized to the 0.1\,--\,1~keV energy band and the corresponding X-ray luminosities are consistent within $1\sigma$ errors except for the HRC-I flux, which seems to be larger. The count rate measured by the HRC instrument is higher by a factor of ca. 2.5 compared to the ACIS-S count rate. The nominal effective areas of the two instruments are very similar at low energies, with the HRC having somewhat larger effective area below 200~eV ($\Delta A \approx 10\ \mbox{cm$^2$}$ or 17\% at 200~eV, 44\% at 150~eV). The additional counts might arise from photons at these energies, but considering the small difference in effective areas, it does not seem likely that this is the case for all excess HRC photons. This mismatch is further validated by comparing ACIS-S and HRC count rates with WebPIMMS: assuming a thermal plasma with solar abundances and $T=1$~MK, 8 counts in the 0.15\,--\,0.65~keV energy band in ACIS-S translate into 9 expected counts in the same energy band in HRC-I, which is obviously inconsistent with the 21 recorded HRC photons only 15 minutes after the ACIS-S observation. The photon count estimate changes by $<20\%$ if one assumes a plasma temperature of 0.8 or 1.25~MK, so a slightly different plasma temperature does not cure the substantial mismatch in the count rates.

Mismatches between HRC and {\em XMM} count rates have been reported before for $\alpha$ Cen \citep{robradeschmittfavata2005, ayresjudgesaar2008}. This mismatch between almost simultaneous HRC and ACIS-S count rates can be explained reasonably by two possibilities: either the effective area of the HRC at low energies is underestimated in the current calibration or the effective areas of {\em XMM} MOS and PN (while using the thick filter) as well as {\em Chandra} ACIS-S are overestimated in that energy range. A detailed cross-calibration effort, preferably with a soft coronal source, could help to resolve any systematic errors in the effective areas of the instruments.

\section{Discussion}

\subsection{Low activity - 51 Peg a Maunder minimum candidate?}

\begin{figure}
\includegraphics[width=0.5\textwidth]{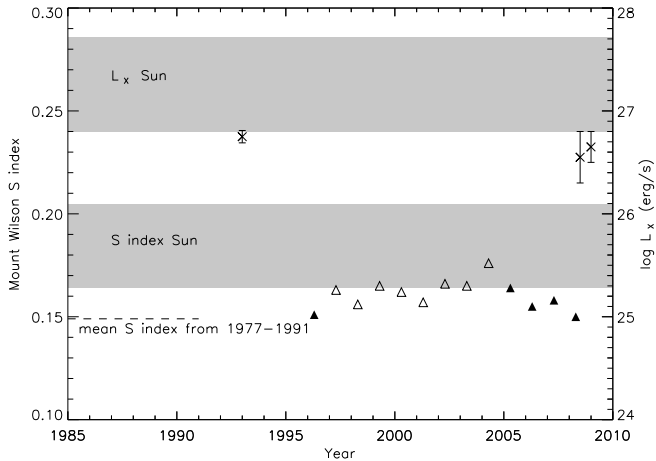}
\caption{X-ray luminosity (crosses) and S index seasonal mean of 51~Peg's \ion{Ca}{ii} H and K line flux; reliable \ion{Ca}{} data given as filled triangles with typical standard deviations of $\lesssim\pm0.005$, open triangles are less reliable values derived from only a small number of observations.}
\label{calcium}
\end{figure}

We found 51~Peg to be a rather constant, weak and soft X-ray source over the last 15~years. Another available long-term activity indicator is \ion{Ca}{ii}. In the \ion{Ca}{ii}~H and K line flux monitoring programs carried out at the Mount Wilson and Lowell Observatories \citep{baliunasdonahuesoon1995, halllockwoodskiff2007}, 51~Peg was found to have a very low chromospheric activity level ($\langle R^{'}_{HK}\rangle=-5.01$, $S_{MW}\approx 0.16$).  In Figure~\ref{calcium} we plot the star's Mount Wilson S~index measured since 1996 together with the average of older data. Clearly, the overall chromospheric activity is low, with some variations in the older set of data which is also seen in the more recent observations. Apart from one data point which is derived from a very small number of observations, 51~Peg's S~indices are at the lower end of or even below the Sun's respective data during a solar minimum (data taken from \cite{baliunasdonahuesoon1995}). Other stellar properties like radius, mass, age and effective temperature are similar to the Sun's respective parameters.

The steady low-activity behavior of 51~Peg's \ion{Ca}{ii}~H and K line fluxes is also reflected by its X-ray properties. Compared to estimates for the solar X-ray luminosity in the {\em ROSAT} RASS band (0.1\,--\,2.4~keV) during a solar cycle \citep{judgesolomonayres2003}, 51~Peg's luminosity is also at the lower end of the Sun's values. The ratio of the star's X-ray to bolometric luminosity is also rather low with $L_X/L_{bol}=1\times10^{-7}$. The X-ray surface flux of F to M~stars was shown to be constrained at the lower end by the surface flux level of a solar coronal hole; $F_{X\,\mbox{(hole)}}\approx10^{4}$~erg~s$^{-1}$~cm$^{-2}$ for the {\em ROSAT} and {\em Chandra} energy band, which translates to $\approx10^{3.8}$~erg~s$^{-1}$~cm$^{-2}$ for {\em XMM}'s 0.2\,--\,12~keV band \citep{schmitt1997}. 51~Peg's surface flux, calculated from the ACIS-S data, is one of the lowest so far detected with $\log F_{X}=10^{3.7}$~erg s$^{-1}$ cm$^{-2}$; the coronal hole surface flux seems to be a good description of this star's X-ray flux, with regards to the flux level as well as the plasma temperature.

There has been some discussion on how to identify a Maunder minimum (MM)~star over the last years. The original criterion of chromospheric activity levels $\langle R^{'}_{HK}\rangle=-5.1$ was derived by \cite{henrysoderblomdonahue1996}, but relied on a stellar sample contaminated with evolved stars, which have significantly lower chromospheric activity levels compared to main sequence stars. \cite{wright2004} reanalyzed these data, excluding evolved stars with luminosities more than one magnitude larger than the {\it Hipparcos} average main sequence for the respective $B-V$ value, and found that most stars previously identified as MM~candidates are evolved stars and therefore not comparable to the Sun's Maunder minimum state. This led \cite{judgesaar2007} to the question if the minimum $\langle R^{'}_{HK}\rangle$ level for main sequence stars to qualify as an MM~candidate should be higher than $-5.1$, and also to consider flat-activity time profiles and UV- and X-ray data to identify MM~candidates. A recent study by \cite{hallhenry2009} suggests that minimum levels of $R^{'}_{HK}$ depend on stellar metallicity, with metal-poor stars from the examined sample having a higher minimal $R^{'}_{HK}$. In this picture, 51~Peg as a metal-rich star still has low chromospheric activity as measured by $R^{'}_{HK}$, but this alone does not necessarily qualify it to be a Maunder minimum candidate. However, as recent results show \citep{halllockwoodskiff2007,hallhenry2009}, the absolute magnetic excess flux $\Delta \mathcal{F}_{HK}$ seems to be a more reliable indicator for stellar activity than $R^{'}_{HK}$. In terms of this quantity, 51~Peg's activity level is even lower compared to the quiescent Sun than indicated by $R^{'}_{HK}$ or the S~index, supporting our interpretation of 51~Peg as being extremely inactive.

The strongest line of evidence for 51~Peg being a Maunder minimum candidate is its flat activity profile as seen over decades in the Mount Wilson program \citep{baliunasdonahuesoon1995} and in observations at Lowell Observatory \citep{halllockwoodskiff2007}, as well as the extremely low X-ray surface fluxes, which have not changed significantly since the 1992 {\em ROSAT} observations. That 51~Peg is a slow rotator with $P_{\star}\approx30-40$~d \citep{baliunassokoloff1996, mayorqueloz1995} fits the picture, making 51~Peg the first MM candidate star with a close-in giant planet.

A statistical analysis of the X-ray luminosities of planet-bearing host stars has recently been conducted \citep{kashyapdrakesaar2008}. Its authors claim that stars with close-in giant planet, such as 51~Peg, are on average X-ray brighter by a factor of two compared to stars with far away planets. Apparently, 51~Peg's overall activity is not enhanced by the presence of its Hot Jupiter. However, at a distance of order of 50~$R_{Jup}$ only a weak interaction between an inactive star and its planet might be expected.

\section{Conclusions}

We have detected X-ray emission from 51~Peg in a 55~ks observation with {\em XMM-Newton} and 5~ks observations with {\em Chandra} ACIS-S and HRC-I each. The detection of 51~Peg with a low count rate in the {\em XMM} pointing and the clear source signal in the much shorter {\em Chandra} observations can be explained by the different effective response of the detectors at low energies and 51~Peg having an extremely cool corona. Our main results are summarized as follows:
\begin{enumerate}
\item 51~Peg shows weak emission in the \ion{O}{vii} triplet and emission around 200~eV which can be explained most likely by cool silicon emission lines.
\item A coronal temperature of $\lesssim 1$~MK is consistent with the detected hardness ratios in different energy bands in both the {\em XMM} and the {\em Chandra} pointing as well as in the {\em ROSAT} observation carried out 16~years earlier.
\item The {\em Chandra} HRC-I count rate is higher than can be explained by differences in the effective areas of HRC and ACIS-S; HRC's effective area might be larger at low energies than given in the calibration so far.
\item The constant and very low surface X-ray flux level together with a flat-activity behavior in chromospheric \ion{Ca}{ii}~H and K line fluxes suggests 51~Peg to be a Hot Jupiter-bearing Maunder minimum candidate.
\end{enumerate}

\begin{acknowledgements}
K.~P.~ and J.~R.~ acknowledge financial support from DLR grants 50OR0703 and 50QR0803. J.~C.~H.~ acknowledges support from grant ATM-0742144 from the National Science Foundation. This work is based on observations obtained with {\em XMM-Newton}, {\em Chandra} and makes use of the {\em ROSAT} Data Archive. We acknowledge the allocation of {\em Chandra} Director's Discretionary Time.
\end{acknowledgements}

\bibliographystyle{aa}
\bibliography{12945}

\end{document}